# NMR quantum information processing

**Dawei Lu[1], Aharon Brodutch[1], Jihyun Park[1], Hemant Katiyar[1], Tomas Jochym-O'Connor[1], and Raymond Laflamme[1, 2, 3]**

[1]*Institute for Quantum Computing and Department of Physics and Astronomy, University of Waterloo, Waterloo, Ontario N2L 3G1, Canada*
[2]*Perimeter Institute for Theoretical Physics, Waterloo, Ontario N2L 2Y5, Canada*
[3]*Canadian Institute for Advanced Research, Toronto, Ontario M5G 1Z8, Canada*

## Introduction

With each passing year computers are used to solve more problems, faster and more efficiently. Nevertheless it seems that many problems are, and will remain, unsolvable by computers based on standard technologies. Three gigantic obstacles, on the road leading to future computers, cannot be circumvented by classical means. (A) Microscopic quantum effects: Moore's law [1] states that the number of transistors in a dense integrated circuit doubles approximately every eighteen months. At this rate in a few years, we will have to store each bit information at the atomic scale, and microscopic quantum effects will play a dominant role. (B) Dissipation: All processes in classical computers are irreversible and dissipative. The energy for implementing a single logical gate is an order of magnitude larger than the Laudaur energy [2], which is the minimal cost consumed by the erasure of a single bit information. This leads to problems of heating and energy consumption that grow with the number of operations per unit time. (C) Computational complexity: Some problems are simply intractable on a classical computer. For instance, storing the state of 71 spins demands $2.4 \times 10^{21}$ bits, approximately the content of all information currently stored by mankind [3]! These limitations require us to seek out radically new technologies for processing and storing information. While the first two problems may be temporarily circumvented by switching to new platforms that implement classical algorithms, there is no doubt that that quantum effects must be dealt with in the near future. Moreover, quantum algorithms are currently the only known way to transverse the obstacle of computational complexity for an important class of problems that include quantum simulations.

Quantum information processors [4] exploit fundamentally new models of computation based on quantum mechanical properties instead of classical physics. While there is no fixed physical platform or the underlying information processing model, most of the known candidates must, almost by definition, deal with quantum effects. Moreover most known models are in principle reversible, and minimize the erasure of information and thermal dissipation [5]. Most importantly there is a good reason to believe that quantum computers can solve some problems exponentially faster than a classical computer [4]. For example, fifty quantum bits (qubits) are sufficient to simulate the dynamical behaviour of fifty spins.

Although fascinating and stimulating, the enthusiasm for building large scale quantum computers [6,7] is partly challenged by the substantial practical difficulty in controlling quantum systems. At present, a number of physical systems can be used to implement small scale quantum processors. These include [6]: trapped ions and neutral atoms, superconducting circuits, spin-based magnetic resonance, impurity spins in solids, photons and others. Amongst all current implementations, nuclear magnetic resonance (NMR) [8,9] has been one of the most successful platforms: having demonstrated universal controls on the largest number of qubits. Meanwhile, many advanced techniques developed in NMR have been adopted to other quantum systems successfully. Therefore, despite the huge difficulties in initialization and scalability, NMR remains indispensable in quantum computing as it continues to provide new ideas, new methods and new techniques, as well as implement quantum computing tasks in this interdisciplinary field.

In Section 1 below we present the basics of quantum information processing (QIP) and the implementation of NMR quantum processors. We show how these processors can satisfy the general requirements of a quantum computer, and describe advanced techniques developed towards this target. In Section 2 we review some recent NMR quantum processor experiments. These experiments include benchmarking protocols, quantum error correction, demonstrations of algorithms exploiting quantum properties, exploring the foundations of quantum mechanics, and quantum simulations. Finally in Section 3 we summarize the concepts and comment on future prospects.

# 1. NMR basics

As of 2015, there are many different proposals for quantum computing architectures and it is unclear which architecture will results in a quantum computer. While the computational model is well defined, the underlying physical implementations are still unknown. There are however, five well-accepted physical requirements [10] that must be satisfied by any potential candidate. The so-called DiVincenzo criteria are: (1) a scalable physical system with well characterized qubits; (2) the ability to initialize the state of the qubits to a simple fiducial state, such as $|000\ldots\rangle$; (3) a universal set of quantum gates; (4) a qubit-specific measurement capability; (5) long relevant decoherence times, much longer than the gate operation time. In this section, we describe how NMR completely or partially satisfies the requirements one by one, and interpret the relevant techniques exploited for each aspect. All concepts, unless specified, refer to liquid-state NMR which is more comprehensible.

## 2.1 Well-defined qubits

A two-level quantum system which is analogous to a spin-1/2 particle, can encode a qubit. The two levels, usually labeled $|0\rangle$ and $|1\rangle$ are the equivalent of $\sigma_z$ +1 and −1 eigenstates respectively. These are often referred to as the computational basis states. Spin-1/2 systems, such as $^1$H, $^{13}$C and $^{19}$F nuclear spins, are natural qubits, and are thus used in vast majority of NMR quantum computation experiments[1]. When a nuclear spin is placed in a static magnetic field $B_0$ along $z$ direction, the dynamical evolution will be dominated by the internal Hamiltonian (set $\hbar=1$)

$$H_\omega = -(1-\sigma)\gamma B_0 I_z = -\frac{1}{2}\omega_0 \begin{bmatrix} 1 & 0 \\ 0 & -1 \end{bmatrix}, \tag{1}$$

where $\gamma$ is the nuclear gyromagnetic ratio, $\sigma$ is the chemical shift arising from the partial shielding of $B_0$ by the electron cloud surrounding the nuclear spin, and $\omega_0 = (1-\sigma)\gamma B_0$ is the Larmor precession frequency. $I_z$ is the angular momentum operator related to Pauli matrix $\sigma_z$ as $\sigma_z = 2I_z$, and so on for $I_x$ and $I_y$. The energy difference between the computational basis states $|0\rangle$ and $|1\rangle$ is the Zeeman splitting $\omega_0$.

For multiple-spin systems, heteronuclear spins are easily distinguished due to the distinct $\gamma$ and thus very different $\omega_0$ in the magnitude of hundreds of MHz, while homonuclear spins are often individually addressed by the distinct $\sigma$ due to different local environments. Furthermore the qubit-qubit interactions are the natural mediated spin-spin interactions called Hamiltonian J-coupling terms. The dipole-dipole interactions are averaged out due to rapid tumbling in liquid solution. The Hamiltonian is

$$H_J = \sum 2\pi J_{ij}\left(I_x^i I_x^j + I_y^i I_y^j + I_z^i I_z^j\right) \approx \sum 2\pi J_{ij} I_z^i I_z^j. \tag{2}$$

The approximation is valid when the weak coupling approximation $\Delta\omega_0 \gg 2\pi|J_{ij}|$ is satisfied, which is always the case for heteronuclear spins and moderately distinct homonuclear spins.

Therefore, the total internal Hamiltonian for a *n*-spin system is

$$H_{int} = -\sum \omega_0^i I_z^i + \sum 2\pi J_{ij} I_z^i I_z^j, \tag{3}$$

which forms a well-defined multi-qubit system used in most NMR quantum computing experiments.

## 2.2 Initialization

Initialization is a process to prepare the system in a known state such as the ground state $|00\ldots\rangle$, which is generally a pure state. The Boltzmann distribution requires extremely low temperatureS, about tens of mK at 1GHz, to prepare such state in liquid state NMR. To avoid working at such low temperatures, a pseudo-pure state

---

[1] For simplicity we only consider spin ½ systems in this chapter.

(PPS) [11,12] is used in almost all NMR experiments.

A *n*-qubit PPS is described as

$$\rho_{PPS} = \frac{1-\varepsilon}{2^n}\hat{I} + \varepsilon|00\ldots\rangle\langle 00\ldots|, \tag{4}$$

where $\hat{I}$ denotes the identity matrix and $\varepsilon \sim 10^{-5}$ is the polarization. In NMR, the identity term is invariant under unital operations; these include pulses, free Hamiltonian evolution, and $T_2$ decoherence. Moreover the identity does not contribute to measured signal as we will see later. Thus the dynamical behaviour of the PPS is the same as that of a pure state. The creation of PPS from thermal equilibrium inevitably involves non-unitary transformations since the eigenvalues of the PPS and thermal equilibrium state are different. Several approaches such as temporal averaging [13], spatial averaging [11], logical labeling [12], and cat-state [14] have been proposed to date. However none of these methods are scalable due to exponential signal decay as a function of the number of qubits [15]. There are scalable methods for preparing qubits in a pseudo-pure state. One such method, algorithmic cooling, is presented in detail in Chapter XXXXXXXXX.

## 2.3 Universal gates

Physically a quantum algorithm is a dynamical process taking the initial state to a final state. One of the breakthroughs of quantum computing was the realization that it possible to efficiently break this dynamical process into a finite set of elementary gates such as the set that includes finite single-qubit rotations and the controlled-NOT (CNOT) gate [4]. In NMR, we can apply external radiofrequency (RF) pulses in the transversal *x-y* plane to realize single-qubit rotations. The external Hamiltonian for a single qubit in the lab frame is written as

$$H_{ext} = -\gamma B_1[\cos(\omega_{rf}t + \varphi)I_x - \sin(\omega_{rf}t + \varphi)I_y], \tag{5}$$

where $B_1$ is the amplitude of the RF pulse, $\omega_{rf}$ is the frequency and $\varphi$ is the phase. For simplicity, we often set $\omega_{rf} = \omega_0$ and work in the rotating frame at frequency $\omega_{rf}$, where the internal Hamiltonian vanishes and the external one remains stationary. Using the propagator in the rotating frame

$$U = e^{i\gamma B_1(\cos\varphi I_x - \sin\varphi I_y)t} \tag{6}$$

and choosing appropriate $B_1$, $\varphi$ and pulse width *t*, one can approximate arbitrary angle rotations around any axis in the *x-y* plane. Using Bloch's theorem we can decompose single-qubit unitary into rotations around two fixed axes

$$e^{i\alpha}R_x(\beta)R_y(\gamma)R_x(\delta). \tag{7}$$

A CNOT gate is a unitary gate with has two input qubits usually called control and target. The gate flips the target when the control is in the $|1\rangle$ state, and does nothing if the control is in the $|0\rangle$ state. In NMR pulse notation the gate can be written as

$$U_{CNOT} = \sqrt{i}R_z^1(\pi/2)R_z^2(-\pi/2)R_x^2(\pi/2)U(1/2J)R_y^2(\pi/2), \qquad (8)$$

where $U(1/2J) = \exp(-i\pi I_z^1 I_z^2)$ indicates the J-coupling evolution for time $t = 1/2J$. The undesired terms in the internal Hamiltonian can be removed via refocusing techniques by inserting π pulses in appropriate positions during the evolution. Note that this refocusing scheme is inefficient to design when the size of system increases. Alternatively, the sequence compiler technique which can track the off-resonance and coupling effects followed with correcting the phase and coupling errors is scalable and particularly powerful [16].

In principle any quantum circuit can be implemented by RF pulses and internal Hamiltonian evolutions. In practice however, the requirement for ultra-high precision is hard to implement. Conventional composite pulses are accurate in small-size systems but hard to scale due to relaxation during the relatively long duration. To overcome this problem, optimized pulse engineering techniques inspired by optimal control theory for NMR quantum computing have been developed in recent years. Here we primarily focus on the GRadient Ascent Pulse Engineering (GRAPE) techniques [17].

*Pulse engineering based on GRAPE algorithm*

Given some theoretical unitary evolution $U_{th}$ the aim of a pulse engineering algorithm is to find an experimental pulse sequence $U_{exp}$ which, together with the free evolution, produces the desired unitary up to some error. The distance between the theoretical and experimental pulse is given by the Hilbert-Schmidt fidelity

$$\Phi = \left|\text{tr}(U_{th}U_{exp}^\dagger)\right|^2/4^n. \qquad (9)$$

In NMR the experimental unitary is an $N$ step digitized pulse where unitary operator for the $m$-th step is

$$U_m = e^{-i[H_{int} + \sum u_k(m)H_{ext}]\Delta t}, \qquad (10)$$

where $u_k(m)$, the controllable RF fields, remains constant at each step. The optimization process starts with a guess for the optimal sequence. At each subsequent iteration we alter $u_k(m)$ according to the gradient

$$u_k(m) \to u_k(m) + \epsilon \frac{\Delta \Phi}{\Delta u_k(m)}. \qquad (11)$$

After a number of iterations the fidelity $\Phi$ will reach a local maximum, and will usually provide a high-fidelity GRAPE pulse to implement $U_{th}$.

The GRAPE optimization method is much faster than conventional numerical optimization methods. It is robust to RF inhomogeneities and drift of chemical shifts, as well as friendly to the spectrometer due to its smoothness. The major drawback of GRAPE technique is the inefficiency with respect to the system size. However, separating the entire system into small subsystems may moderately reduce the complexity [16]. Another drawback is possible discrepancies between the designed pulse and implemented pulse. A feedback system called pulse fixing can be employed to correct these systematic imperfections.

*Pulse fixing*

Non-linearities in pulse generation and amplification, and bandwidth constraints of the probe-resonant circuit, prohibit a perfect match between the designed pulse and real one. The solution is measuring the control field at the sample and closing a feedback loop which can iteratively adjust the control pules so that the real field at the sample matches the designed one. First a pickup coil is used to measure the fields in the vicinity of the sample, and the data is fed back to compare with the target pulse. Then a new pulse attempting to compensate the imperfection is generated based on the measurement result, and sent back to the pulse generator. A good match between design and experiment is typically reached after 8-10 loops.

## 2.4 Measurement

The measurement in NMR is accomplished with the aid of an RF coil positioned at the sample. This apparatus can detect the transversal magnetization of the ensemble, and transform the time-domain signals into frequency-domain NMR spectra via Fourier transform. The detection coil is very weak coupled to the nuclear spins, and does not contribute much to decoherence. However, due to the interactions with the heat bath and inhomogeneity of the static field, the nuclear spins still decohere, leading to free induction decay (FID) of the time-domain signal. The weak measurement process cannot extract much information from a single spin and is not projective. Nonetheless, the ensemble averaged measurement provided by bulk identical spins can, for some purposes, provide more information than a projective measurement.

The FID measurement allows us to extract the expectation values of the readout operators in the *x-y* plane in a single experiment. Pauli observables outside the *x-y* plane can be rotated into the *x-y* plane first and then measured in the allowed basis. In this way full quantum state tomography [18,19] is achievable in NMR to determine all elements of the density matrix describing the quantum state.

## 2.5 Decoherence

Decoherence remains a fundamental concern in quantum computation as it leads to the loss of quantum information. It is traditionally parameterized by the energy relaxation rate $T_1$ and the phase randomization rate $T_2$. $T_1$ originates from couplings between the spins and the lattice, which are usually tens of seconds in an elaborate liquid sample. $T_2$ originates from spin-spin interactions such as the unaccounted terms in the internal Hamiltonian. For NMR quantum computing the timescale $T_2^*$ which involves the effect of inhomogeneous fields is often more important than the intrinsic $T_2$ time. Characteristic $T_2^*$ ranges from tens of milliseconds to several seconds, compared to the two-qubit gate time about several milliseconds. For simple quantum tasks, it is sufficient as hundreds of gates can be finished before $T_2^*$ has elapsed. However, for complex algorithms, other ideas have to be employed to counteract the decoherence and preserve the information. Here we introduce how to use RF selection technique to improve $T_2^*$.

*RF selection*

In NMR, the RF inhomogeneity can be mostly eliminated by running a RF selection sequence. This is a sequence of pulses followed by a gradient field that removes polarization on all but a small part of the sample. The part which is left polarized is restricted to a more homogeneous field strength. We have used the RF selection sequence as

$$R_x(\pi/2)[R_{-x}(\pi)]^{64}\left[R_{\theta_i}(\pi)R_{-\theta_i}(\pi)\right]^{64}R_y(\pi/2) + Gradient, \qquad (12)$$

where the sum over $\theta_i$ should be $\pi/8$ and the number of loops can be varied according to the requirement of homogeneity. The RF selection sequence is applied prior to computation sequences in the experiment, and the signals produced by the inhomogeneous portion of the sample are discarded after the gradient pulse. RF selection improves the timescale $T_2^*$ significantly at the cost of signal loss. For instance, rising the homogeneity to ±2% will result in around 12% residual signal.

## 2. Recent experiments

### 2.1 Benchmarking

Characterizing the level of coherent control is important in evaluating quantum devices. It allows for a comparison between different devices, and indicates the prospects of these devices with respect to the fault-tolerant quantum computing [20]. The traditional approach for characterizing any quantum process is known as quantum process tomography (QPT) [21,22], which has been realized in up to 3-qubit systems in experiment [23–28]. However an arbitrary process on a *n*-qubit system has $O(2^{4n})$ free parameters. So, while QPT fully characterizes the process it requires exponential number of experiments, making it impractical even for moderately large systems. For most practical purposes however, we do not need to determine the value of all free parameters experimentally. For quantum error correction, a few parameters related to the level of noise are required. Several useful techniques such as twirling [29–31], randomized benchmarking [32–34], and Monte Carlo estimations [35,36] can be used to characterize a given quantum process. In the following we describe an experimental realization of the twirling process.

Twirling is the process of conjugating a quantum process by random (Harr-distributed) unitaries. The quantum process is then reduced to a depolarizing channel with a single parameter to describe the strength of the noise. The twirling procedure provides a way to estimate the average fidelity of an identity operation with a few experiments depending only on the desired accuracy. It can be extended to characterize the noise of various unitary operations such as those in the Clifford group[2]. It is however, not useful for separating preparation and measurement errors. Randomized benchmarking is the generalization of twirling by applying a sequence of Clifford gates and measuring the fidelity decay as the function of an increasing number of gates. The decay rate is independent of the preparation and readout errors up to an additional normalization, but the shortcoming is it cannot provide the information for a particular quantum process. Monte Carlo estimations have the same scaling as the twirling protocol in the case of Clifford gates, but are not as natural as twirling if the probability of a given weighted error is required.

In this section, we focus on the twirling and randomized benchmarking protocols, and describe the relevant NMR experiments [29,31,34,37] briefly. We show that reliable coherent control has been achieved in NMR quantum computing up to seven qubits.

*Characterization of a quantum memory*

Ideally qubits in a quantum memory do not evolve dynamically, i.e. the dynamical evolution is the identity. The original twirling protocol, proposed by Emerson *et al*. [29], considers noisy quantum memories, where the quantum process is a faulty identity $\Lambda$. The average fidelity of this process is defined as

$$\bar{F}(\Lambda) = \int d\mu(\psi) \langle\psi|\Lambda(|\psi\rangle\langle\psi|)|\psi\rangle, \tag{13}$$

where $d\mu(\psi)$ is the unitary invariant distribution of pure states known as Fubini-Study measure [32]. It is equivalent to average over random unitaries distributed according to the Harr measure $d\mu(V)$ [32].

$$\bar{F}(\Lambda) = \int d\mu(V) \langle\psi|V^\dagger \circ \Lambda \circ V(|\psi\rangle\langle\psi|)|\psi\rangle, \tag{14}$$

Formally the continuous integral of Eq (14) can be replaced by a finite sum over a unitary 2-design [30] such as, the Clifford group C.

$$\bar{F}(\Lambda) = \frac{1}{|C|}\sum_{C_i \in C}\langle\psi|C_i^\dagger \circ \Lambda \circ C_i(|\psi\rangle\langle\psi|)|\psi\rangle. \tag{15}$$

It is possible to show [29] that the average fidelity above can be derived from an experiment where instead of conjugating by elements of the Clifford group C we conjugate over the single-qubit Clifford group $C_1$ applied to each qubit individually together with a permutation. In this way $\Lambda$ is symmetrized to a Pauli channel instead of a depolarizing channel, however the task of noise characterization is simplified to the finding the probability of no error Pr(0) in this channel.

---

[2] The Clifford group is the group of unitary operations that leave take Pauli operators to Pauli operators.

$$\bar{F}(\Lambda) = \frac{2^n \Pr(0) + 1}{2^n + 1}, \qquad (16)$$

where $n$ is number of qubits.

To measure the probability of no error Pr(0), we can probe the Pauli channel with the input state $|00\ldots0\rangle$, followed by a projective measurement in the $n$-bit string basis. For an ensemble system such as NMR, we can use $n$ distinct input states $\rho_w = Z_w I_{n-w}$, where $w$ is the Pauli weight defined as the number of nonidentity factors in the Pauli operator. This is followed by a random permutation operation $\pi_n$ and a random 1-qubit Clifford. The average of the output states returns the scaled parameter of the input. Fig. 1 shows the circuit for both cases.

To estimate Pr(0) to precision $\delta$ it is enough to perform $\log(2n/\delta^2)$ experiments [29], such that each experiment requires random conjugation by a 1-qubit Clifford and permutation.

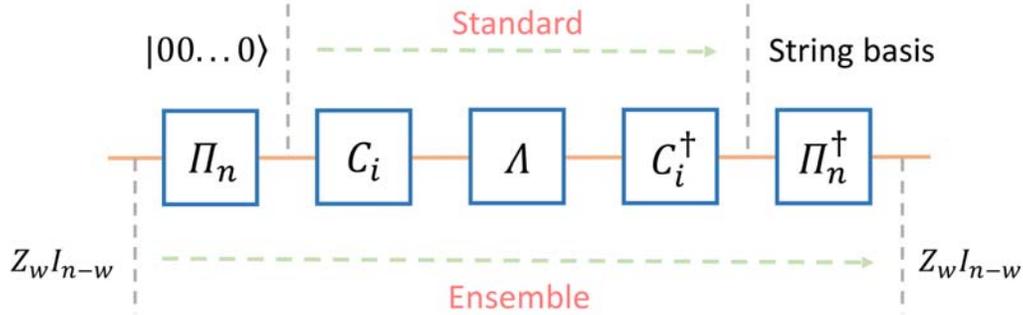

Fig. 1 Quantum circuit to implement the twirling protocol for the purpose of quantum memories. The standard one requires one input state $|00\ldots0\rangle$ and conjugation by $C_i$, whereas the ensemble one requires $n$ distinct input states and an additional permutation $\pi_n$.

The NMR experiments to demonstrate the above protocol were implemented on both a 2-qubit liquid sample chloroform CHCl$_3$ and 3-qubit single-crystal sample Malonic acid C$_3$H$_4$O$_4$. To evaluate the level of quantum memories, C48 pulse sequence [38] was utilized to suppress the evolution of the internal Hamiltonian. Two experiments including one cycle of a C48 sequence with 10μs pulse spacing and two cycles C48 with 5μs pulse spacing were performed to characterize the unknown residual noise. The results show that the probabilities of one-, two-, and three-body noise terms all decrease substantially by using the latter sequence. For instance, Pr(0) increases from 0.44 for the one-cycle case to 0.84 for the two-cycle case.

*Characterization of Clifford gates*

Moussa *et al.* proposed [31] a slight modification of the original twirling protocol, to efficiently estimate the average fidelity of a Clifford gate, by inserting an identity process in the original twirling protocol. If a noisy quantum operation $U_N = U \circ \Lambda$ can be represented by a noisy process $\Lambda$ followed by the application of the target unitary $U$, an identity process written as $U^\dagger \circ U$ can be inserted between $C_i^\dagger$ and $\Lambda$ in Eq (15). Thus the average fidelity of a noisy identity $\Lambda$ transforms to the average fidelity of a noisy unitary gate $U_N$

$$\bar{F}(\Lambda) = \frac{1}{|C|} \sum_{C_i \in C} \langle \psi | C_i^\dagger U^\dagger \circ U_N \circ C_i(|\psi\rangle\langle\psi|) | \psi \rangle. \tag{17}$$

This is further simplified by combining the pieces following $U_N$ into a new measurement $M_{new} = U C_i M C_i^\dagger U^\dagger$. Note that for an ensemble system the permutation operations $\pi_n$ need to be applied accordingly, see Fig. 2. In general, this new measurement is difficult to implement as $U$ is usually impossible to realize in experiment. However, if $U$ is an element of the Clifford group and the original measurement $M$ is a Pauli measurement (always the case in NMR), $M_{new}$ can be calculated efficiently. The remaining steps are the same as the characterization of quantum memories described in the original twirling protocol.

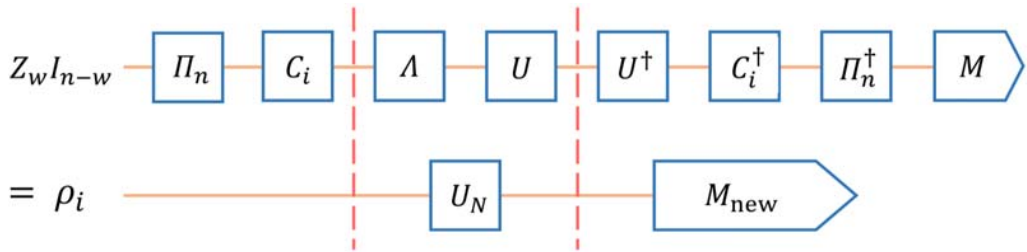

Fig. 2 Quantum circuit for modified twirling protocol to certify noisy Clifford gates. An identity process $U^\dagger U$ is inserted to generate the aim $U_N$. $\rho_i = C_i \pi_n (Z_w I_{n-w}) \pi_n^\dagger C_i^\dagger$ is a random Pauli state and $M_{new} = U C_i \pi_n M \pi_n^\dagger C_i^\dagger U^\dagger$ is an efficiently pre-computed Pauli operator if $U$ is a Clifford gate.

Despite the limitation of the protocol to characterize only Clifford gates, the modified twirling protocol is still significant since Clifford gates construct the elementary units in the vast majority of fault-tolerant quantum computations based on stabilizer codes, where universality is granted by magic state preparations [39]. In addition, the evolution of states under Clifford gates can be tracked efficiently, as mentioned above.

The first experiment for certifying Clifford gates was implemented by a 3-qubit single-crystal sample Malonic acid $C_3H_4O_4$ in NMR. The aim was to certify the encoding operation of the phase variant of the 3-qubit quantum error correcting code [40]. This can be decomposed into two CNOT gates and three single qubit Hadamard gates. A GRAPE pulse [17] with the length 1.5ms was designed to implement this encoding operation, and was rectified by pulse fixing in experiment. The estimation of the average fidelity before and after the rectification is 86.3% and

97.3%, respectively. After factoring out the preparation and measurement errors, the average fidelity of the rectified implementation was improved to be over 99%.

Another experiment of certifying a 7-qubit Clifford gate in NMR was carried out recently [37]. The sample was Dichlorocyclobutanone, and the seven carbons formed a 7-qubit quantum processor. The target Clifford gate was chosen as the one to generate the maximal coherence from single coherence with the aid of a few local gates. A 80ms GRAPE pulse was obtained to realize this operation with a theoretical fidelity over 99%, followed with the rectification by pulse fixing in experiment. 1656 Pauli input states were randomly sampled out of the entire Pauli group, which consists of 16383 elements, to achieve a 99% confidence level. The average fidelity of this Clifford gate is 55.1%, which is reasonable because high weight Pauli states are extremely fragile to the effect of decoherence. To assess the gate imperfections, the decoherence effect was simulated under the assumption that it could be factored out. The average fidelity with the elimination of decoherence increased to 87.5%. As the Clifford gate involved about six two-qubit gates and twelve single-qubit gates, the average error per gate was estimated about 0.7%, attributed to the imperfections in designing and implementing the GRAPE pulse. The NMR spectra observed after applying this Clifford gate were used as further evidence of the level of control.

*Randomized benchmarking of single- and multi-qubit control*

The twirling protocol requires a lower error rate in preparation and measurement than the quantum gate being certified, and cannot identify errors that are due to preparation and measurement. As a result, randomized benchmarking was developed as a modification of the twirling protocol that enables the estimation of error rates per gate for a particular quantum system, independent of the preparation and measurement errors.

The idea is similar to the one used for characterizing a single Clifford gate, but replacing the single Clifford gate with a sequence of one- and two-qubit Clifford gates, that are uniformly sampled from the Clifford group. The outcome is the fidelity decay as a function of the number of Clifford gates in the sequence. Assuming that the errors are independent of the gates, the preparation and measurement errors provide only an additional normalization to the fidelity decay curve. Note that the sequence must be constructed by Clifford gates, to enable the output state tracking and reversal gate designing.

Both single- and three-qubit experiments have been performed in NMR [34]. For the single-qubit experiment, the sample was unlabeled chloroform. The Clifford gates were randomly chosen from $\pi/2$ and $\pi$ rotations around $x$, $y$, or $z$ axis, and applied sequentially for utmost 190 times. Traditional Gaussian pulses, BB1 composite pulses and GRAPE pulses were all tested, with the average fidelity $2.1 \times 10^{-4}$, $1.3 \times 10^{-4}$, and $1.8 \times 10^{-4}$, respectively. The results were initially somewhat surprising as the optimized GRAPE pulses could not surpass the performance of the BB1 pulses. The explanation is that GRAPE pulses are more sensitive to implementation imperfections such as finite bandwidth effects.

3-qubit experiments were performed using $^{13}$C-labeled trissilane-acetylene, and the Clifford group generating set was chosen to be the Hadamard, PHP$^\dagger$(a Hadamard conjugated by a phase gate) and nearest-neighbour CNOT gates. The sequence was constructed by randomly choosing the gates from the above group, with 2/3 probability to implement the single-qubit gates and 1/3 probability to implement the two-qubit CNOT gates. All the operation were optimized by 99.95% fidelity GRAPE pulses. Starting from a fixed initial state *ZII*, the theoretical output could be tracked and recovered to return to *ZII* in the end. By comparing the signal loss as a function of number of gates and fitting the exponential fidelity decay, the average error per gate is about $4.7 \times 10^{-3}$, which is an order of magnitude higher than the expected error $4.4 \times 10^{-4}$ obtained from the GRAPE imperfection and decoherence. This implies there are still unknown factors which have yet to be handled in the pulse design.

## 2.2 Error correction and topological quantum computing

To build a reliable and efficient QIP device, a quantum system should be resilient to errors caused by unwanted environmental interaction and by imperfect quantum control. The progressive development of quantum error correction codes (QECC) and fault tolerant methods in the past two decades have been central to determining the feasibility of implementing a quantum computer. The threshold theorem proves that implementing a robust quantum computer is possible in principle, provided that the error correction schemes can be implemented physically above a certain accuracy [41,42,43]. Using error correction methods, a quantum computer can tolerate faults below a given threshold that depends on the error correction scheme used. Despite many outstanding achievements in the theoretical field, implementing such schemes in physical experiments remains a significant challenge, in large due to the requirement for a relatively large number of qubits. NMR was one of the early platforms used to take up the challenge of demonstrating fault tolerance in real experiments. In this section, we outline some of the fundamentals of fault tolerance and review some experimental implementations of QECC in NMR.

The basic ideas behind quantum error correction schemes [4,47] are similar to those used in classical error correction methods, which exploit the idea of redundancy. Suppose the information carried in a classical bit is copied (encoded) onto two other bits. The *logical* bit is now the encoded three *physical* bits. For simplicity, let us assume we are interested only in storing the information. During the storage time, bit flipping errors can occur, and the errors need to be corrected. For this purpose, we need to check repeatedly whether the three bits are all in the same state. Whenever a discrepancy occurs, we can use a majority vote to bring all three bits to the same states. The method assumes that the probability of two erroneous bits is much lower than the probability of a single erroneous bit.

Extending the classical scheme to the quantum case requires some care. First of all, we need to avoid direct measurement on an encoded state to prevent a quantum superposition from collapsing. As shown in Fig. 3, this can be done by adding ancilla qubits to the circuit and applying appropriate gates to obtain sufficient information about the error without learning the state of the logical qubit. Subsequently, the ancilla qubits can be measured to obtain an error syndrome, which ideally contains the relevant information about which qubit is affected by what kind of error. Secondly, a quantum encoding scheme can exists without violating the 'no cloning principle' of quantum states. The quantum encoding step repeats the state only on a computational basis, which is different from copying a state (i.e. $\alpha|0\rangle + \beta|1\rangle \xrightarrow{encoding} \alpha|000\rangle + \beta|111\rangle$, not $(\alpha|0\rangle + \beta|1\rangle)^{\otimes 3}$).

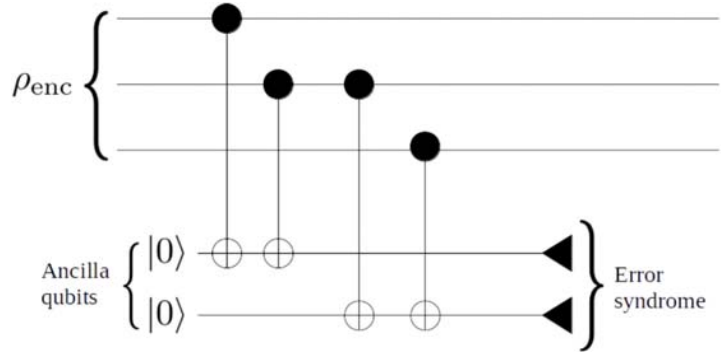

Fig. 3 To detect the error in the encoded state, the ancilla qubits are used to obtain information about the error. Subsequently, the error syndromes can be used to detect and identify the errors. With this knowledge, we can recover the state by applying appropriate recovery operation to the erroneous state.

Any decoherence phenomenon can be decomposed into two types of errors: X and Z. Here, X and Z are Pauli matrices. X (or bit flip) errors flip the spin states, from $|0\rangle$ to $|1\rangle$ and vice versa, whereas Z (or phase flip) errors have the effect

$$\frac{1}{\sqrt{2}}[|0\rangle + |1\rangle] \underset{Z}{\Leftrightarrow} \frac{1}{\sqrt{2}}[|0\rangle - |1\rangle]. \tag{18}$$

In a NMR system, Z errors are similar to the $T_2$ decoherence effect; while $T_1$ is in the class of X errors although it is not symmetric. Depending on circumstances, a code might focus on correcting one of the two types. Since the two types are closely related, a code that corrects one type can be easily modified to fix the other type. To correct for both types of errors, a larger code might be necessary. For example, Shor's nine qubit code [44] is a code based on two 3-qubit codes, one for X errors and one for Z errors.

Several early quantum error correction codes, such as the three qubit phase error correcting code and five qubit error correcting code, were first implemented using a NMR QIP device [45–47]. The three qubit phase error and five qubit error correction codes correct a single qubit phase error and an arbitrary single qubit error, respectively [48]. These implementations demonstrated the benefits of quantum error correction, showing that error correction indeed protects the quantum information even in the presence of gate imperfections. Ideally, to protect quantum information from decoherence, the error corrections should be applied repeatedly. This requires resetting ancilla qubits to a pure state after each error correction state, such that they can be reused for the next round. To achieve this, the time it takes to reset ancilla qubits should be considerably shorter than the duration of a desired circuit. In NMR the natural reset time is $T_1$. Consequently, the lifetime of the circuit is comparable to the time it takes to reset the ancilla qubits. Moreover, even if the $T_1$ values of ancilla qubits are an order of magnitude shorter, these qubits will reset to a thermal equilibrium state (in NMR systems) instead of the desired pure state. Therefore, existing NMR implementations are limited to a single round of correction. A single round includes encoding, decoherence, decoding and error correction steps. The solution to this problem is an active area of research. Methods such as algorithmic cooling can be used to increase the number of rounds [49].

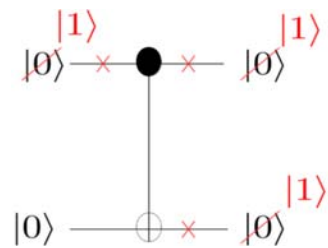

Fig. 4 A bad propagation of errors. An error in the control qubit propagates to the target qubit as well.

Recently, Zhang et al. [50] realized the implementation of logical gates (Identity, Not, and Hadamard) on encoded qubits to demonstrate the use of quantum error correcting codes in an information processing task. The previous demonstration of single round of the five qubit error correction code [47] was extended by applying a logical gate after the encoding step. To realize this additional step, they used a dipolar coupled system which reduced the duration of the experiment by the order of magnitude compared to Ref. [47], which was about 300ms. To implement each gate, each of the five physical qubits was subjected to I, X, Z, and XZ errors (16 possible errors). Comparing the averaged fidelities of the 16 outgoing states with and without the error correction, they showed that the gates perform better when error correction is applied. The improvement in terms of average fidelity was 0.0837, 0.0528, and 0.0196, for Identity, Not and Hadamard gates respectively.

To perform fault-tolerant quantum computing, it is important that errors do not propagate badly. It is critical to ensure that when qubits interact, errors from one qubit do not propagate uncontrollably to the rest of the system. One way to do so is to compute and correct errors transversally [51,52]. For example, consider applying a CNOT gate between two logical qubits (see Fig. 4), a bit flip error on the control qubit will propagate to the target qubit[3].

In an error correction model, different methods to implement the logical gates can lead to different models for error propagation. For example, there are two ways to apply a CNOT between two logical qubits, encoded using the three-qubit code, $\alpha|0\rangle + \beta|1\rangle \rightarrow \alpha|000\rangle + \beta|111\rangle$ (see Fig. 5). In the first case, an error occurs on the first qubit of the first block. This error will propagate to all the qubits in the second block, causing a logical error which is not correctable. By implementing a CNOT gate as shown in the second case, we can avoid this bad propagation of errors. One error on the logical control qubit will propagate to at-most one error in the logical target qubit. Each can be corrected individually. This second case is an example of a transversal CNOT gate.

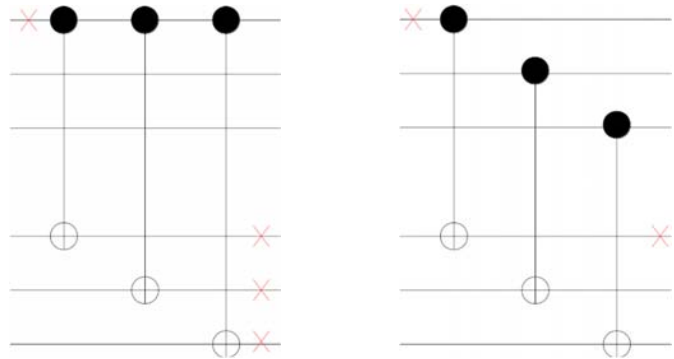

Fig. 5 Two different ways to implement a CNOT gate between the two three-qubit codes: the example on the left is an example of a bad implementation of a CNOT gate between the two three-qubit codes. An error on the first qubit of the first block can propagate to all the qubits in the second block. On the contrary, the right example propagate to at most one error on the bottom encoded qubit and thus remains correctable.

By applying the gates transversally between the encoded blocks, we can ensure errors propagate in a controllable manner. Unfortunately, for any given code, it is not possible to implement a universal set of gates transversally [53]. One way to overcome this problem in constructing a universal fault-tolerant quantum computer is to use special states known as magic states along with the transversal gates [54–56,60]. These special states can be used together with transversal gates to perform an operation which is otherwise not transversal. Other proposed techniques include concatenation of two different codes [57], transversal gates with the addition of gauge fixing [58], as well as code conversion [59]. However, magic state preparation is currently one of the most well-studied techniques. One method of special interest is the transversal Clifford plus the use of the T-type magic state used to implement the remaining gate in the universal gate set [39].

---

[3] In general, a protected space does not necessarily encode a single qubit, and single qubit gates are often not transversal.

Preparation of a high fidelity magic states may be critical for universal fault-tolerant quantum computation. *Magic state distillation* is a method to prepare a number of high fidelity magic states from the larger number of low fidelity magic states. Souza et al. [61] demonstrated magic state distillation for the first time using NMR. They showed that they have sufficient control to distill five imperfect magic states into a single higher fidelity magic state. The higher fidelity target T-type magic state

$$\rho_M = [I + (\sigma_x + \sigma_y + \sigma_z)/\sqrt{3}]/2 \qquad (19)$$

was prepared from the five copies of imperfect states

$$\rho = [I + p(\sigma_x + \sigma_y + \sigma_z)/\sqrt{3}]/2 \qquad (20)$$

where $p = \sin \alpha$, $\alpha \in [\pi, 3/2\pi]$. To quantify how close the imperfect state is from the target magic state, they measured the m-polarization defined as $p_m = 2\text{Tr}(\rho_M \rho)$. They showed that the m-polarization increases after performing the magic state distillation based on the five qubit error correction code, if the input m-polarization (m-polarization averaged over the five imperfect states) is large enough (>~0.65).

*Topological Quantum Computing (TQC) using anyons*

If the quantum error correction method introduced above is an algorithmic way to protect quantum information, topological quantum computation is the work towards realizing a physical medium that is naturally resilient to decoherence. Anyons, exotic quasiparticles, can be used to realize such a medium. In anyonic topological codes, the computation is encoded in a degenerate ground state of a two dimensional system that supports anyons [62].

Due to their fault-tolerant nature, anyonic systems have been gaining much attention lately for their prospects in building quantum memories and topological quantum error correction architectures [62,63]. Currently, most of the contributions to the field have been theoretical, due to the difficulty in building the relevant topological systems. Here we will briefly discuss the past and future contributions of NMR to the field of experimental topological quantum computation. To lead to this discussion, we will first introduce anyons and their properties. For simplicity, we will treat anyons as fundamental particles, although a more correct term would be quasi-particles, excitations that behave like particles. After discussing the basic properties, we will discuss how fault-tolerant quantum computation can be achieved with anyons in a general setting, and we shall conclude the topic with an explanation of a specific anyonic error correction scheme which has been demonstrated in NMR.

Anyons are particles that can be created in a two dimensional system, such as a spin lattice system [64]. They have a unique property which distinguishes them from other fundamental particles such as bosons or fermions. This property, known as fractional statistics, plays a key role in TQC. Unlike bosons and fermions, anyons behave in a non-trivial way under particle exchange.

This operation of exchanging the positions of particles is referred to as a 'braiding operation' [62] in general. Here, we will focus on an operation where we exchange the positions of two particles twice. Intuitively, we would expect the system to come back to its original state as in the case of bosons and fermions. However, if we exchange the positions of two anyons twice, the wavefunction can either obtain a phase factor, ranging from $0$ to $2\pi$ (abelian anyon), or evolve according to a unitary matrix (non-abelian anyon).

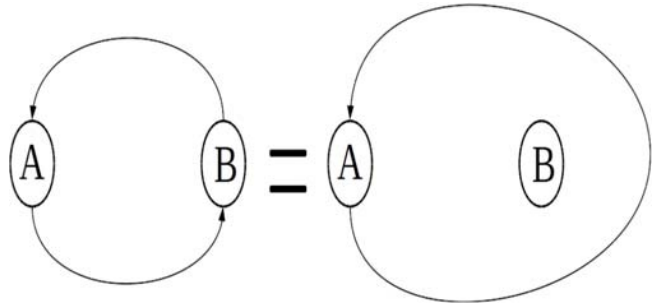

Fig. 6 The operation that exchanges the particles twice is equivalent to circulating one particle around the other. The effect of this operation in two dimensions depends on the topology.

This interesting phenomenon is the result of different topologies that can be manifested in the two dimensional case. Exchanging the positions of two particles twice is equivalent to moving one particle around the other as shown in Fig. 6. Such a braiding path can always be contracted to a point for a three dimensional case, whereas the braiding path confined on a plane (a two dimensional case) cannot be contracted to a point (for more detailed explanation, refer to Fig. 7). This may result in anyons with non-trivial statistics.

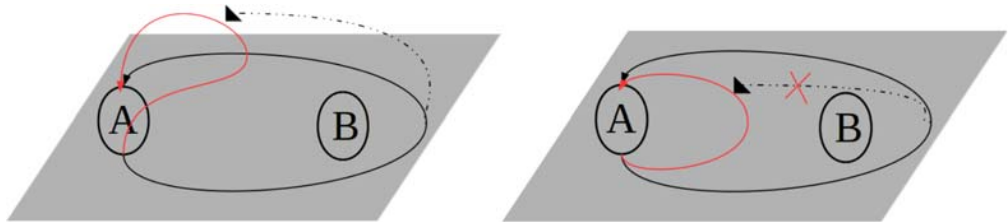

Fig. 7 The topological difference of the braiding path between the three dimensional (left) and the two dimensional (right) cases (circles with A and B represent particles): the braiding path (black), can be smoothly deformed to the red path for the three dimensional case (imagine the situation where we are pulling one end of the black loop, and we also have freedom to move this loop up and down). Similarly, this red loop can be further deformed to a point, which is effectively doing nothing on the system. However, in a two dimensional case, since the black loop is confined in a plane, we cannot continuously deform the path to the red one (unless we make a cut). Here, again, imagine pulling one end of the black loop, but without the freedom of moving the loop up and down. The loop gets stuck because of the particle B.

Although we cannot build a truly two-dimensional system [62,63], we can physically realize an effective two dimensional system. Therefore, anyons do not appear as fundamental particles, but as quasiparticles, usually localized defects or excitations of quantum systems.

The quantum states of anyons can be used for QIP. By creating and moving anyons around, we can encode our information in the anyonic state space. As long as the quasi-particles are stable, the information stays safe. Gate operations can be implemented by braiding anyons. In this way, we ensure that the state evolves exactly, since the statistics constitute a unique particle property. This property is exact and path independent. It does not matter which paths we use to braid the anyons, as long as the topology of the paths are the same. In other words, the path depends only on its global (topological) property, not a local property. This feature provides flexibility when implementing a braiding operation, since there is no need to form a precise loop and small wiggles in the path have no effect.

To use anyons for quantum computing, we should first come up with a consistent mathematical model which describes the braiding and fusion statistics of different types of anyons [62,65]. Working out a mathematical model with appropriate braiding and fusion statistics to implement a desired circuit is non-trivial. Moreover, having a mathematical anyonic model which can perform universal set of quantum gates using only topological operations, i.e. the Fibonacci anyonic (non-abelian) model [62], does not necessarily mean we can realize the model experimentally. In fact, although there has been experimental demonstration of existence of abelian anyons [66], the experimental evidence of non-ablien anyons is still not conclusive, despite extensive ongoing progress [67].

While topological quantum computing is fascinating, there is a lot of interest in using anyons for only part of the design, for example in quantum memories and quantum error correction codes. Since many of these ideas rely on exploiting abelian anyons, they can be physically realized with near future technology.

One well-known topological quantum error correction scheme is the quantum double model [62,63]. In this model, quantum information is encoded into a collective state of interacting spins on a two-dimensional surface. This encoding scheme takes the spatial relationship between the qubits into account, unlike conventional QECC. These topological systems have certain important properties. First of all, the Hamiltonian of these systems possesses a finite energy gap between degenerate ground states and the excited states. Information encoded in the ground state is therefore protected by this energy gap, since a jump from the ground state to the excited state has an associated energy cost. Also, the information is encoded in a non-local way, for example, not only in one particular spin but rather in the state of the entire system. Hence, local errors cannot alter the information. Most importantly, any local errors (excitations) are realized as the creation of a pair of anyons. This is the critical underlying feature which give rise to the properties mentioned above. Thus, the errors can be corrected by annihilating the anyons, connecting the two through a topologically trivial loop which is contractible to a point. Some gate operations can be realized by moving anyons through topologically non-trivial loops. Such encoding schemes have several desirable features: high feasibility, requiring only nearest-neighbor interaction; robustness to local perturbations; and access to topological operations.

Since the underlying Hamiltonian that supports anyons is not the natural Hamiltonian of a molecule in a magnetic field, anyonic systems are not directly implementable in a NMR system. However, there has been an experimental demonstration of the toric code [63], an example of quantum double models. Feng et al. [68] took a state preparation approach. Instead of realizing the Hamiltonian of the toric code, they prepared its ground state, which is a highly entangled state. This approach cannot realize all the properties of the toric code, notably the protection of the ground states by the energy gap. However, it can be used to study the properties of the codespace (the grounds state) that are independent of the Hamiltonian. Particularly, how those properties behave under non-ideal noise and whether we have sufficient control to realize the braiding operations on the codespace with current technology. With the state preparation approach, Feng et al. [68] simulated a small instance (six qubit system) of the toric code [69] and demonstrated operations equivalent to the creation, manipulation and braiding operations of anyons in the toric code system. The experiment showed that we have a sufficient control to realize such operations. Similar experiments were also performed in quantum optics [70,71]. Extending such experiments, we can also explore the path independence property of anyonic braiding operations with NMR. Such small-system experiments of NMR QIP make small steps towards experimental TQC.

## 2.3 DQC1

While liquid state NMR was the first test bed for quantum information processing, objections about the 'quantumness' of this platform were raised early on due to the amount of noise in the system. One objection [72] was that liquid state NMR systems at room temperature could not produce entanglement - Schrodinger's "characteristic trait of quantum mechanics" [73] - and are therefore not quantum in the real sense of the word. To challenge that idea, Knill and Laflamme [74,75] came up with an algorithm which is specifically tailored to NMR systems. The algorithm called *Deterministic quantum computing with 1 qubit* (DQC1), was designed for a processor that has highly mixed input states, intermediate unitary operations and ensemble readout. The input state has *n* spins in the maximally mixed state and one pseudo-pure spin.

For simplicity we describe a specific version of the DQC1 algorithm for estimating the normalized trace of a $2^n \times 2^n$ unitary matrix. This task is expected to be hard to compute (*i.e.* it scales exponentially with *n*) due to the exponential number of elements in the sum[4]. Consider the circuit in Fig. 8, the input is the $n+1$ qubit state,

$$\frac{1}{2^{n+1}}[2\epsilon|0\rangle\langle 0| + (1-\epsilon)I] \otimes I^{\otimes n}. \tag{21}$$

We call the first (pseudo-pure) qubit the *control* and the other $n$ maximally mixed qubits the *target*. The output is

$$\frac{1}{2^{n+1}}\left[2\epsilon|0\rangle\langle 0| \otimes I^{\otimes n}\right] + \frac{1-\epsilon}{2^{n+1}}[|1\rangle\langle 1| \otimes U_n I^{\otimes n} U_n^\dagger \tag{22}$$

$$+ |1\rangle\langle 0| \otimes U_n I^{\otimes n} + |0\rangle\langle 1| \otimes I^{\otimes n} U_n^\dagger].$$

Tracing out the target we get the final state for the control,

$$\rho_f^c = \frac{1}{2}I + \frac{1}{2^{n+1}}\text{Re}[\text{Tr}(U_n)]\sigma_x + \frac{1}{2^{n+1}}\text{Im}[\text{Tr}(U_n)]\sigma_y. \tag{23}$$

The readout $\langle\sigma_x\rangle$ and $\langle\sigma_y\rangle$ is natural for NMR. We assume that $U$ has an efficient description in terms of a quantum circuit so that the algorithm is relatively easy to implement.

To prepare the initial state, it is possible to use the thermal state and depolarize the $n$ target qubits. One way to achieve this is to rotate them into the *x-y* plane, and then apply a gradient to the magnetic field. The gradient randomizes the phase and the target system is left in a maximally mixed state.

The DQC1 computational complexity class is the class of problems that can be solved efficiently using the DQC1 model. This class is unchanged even if we allow a constant k pseudo-pure qubits (DQC-k), instead of a single pseudo-pure qubit. To show that a processor has the computing power to solve problems in this class, it is sufficient to show that it can solve a single problem which is complete for this class[5]. The first experimental implementation of a complete problem for DQC1 was an algorithm for approximating the Jones polynomial at the fifth root of unity [76,77], which is a problem derived from knot theory. Details of the problem are given in Ref. [77], for our purposes it is sufficient to note that the algorithm is a special case of the algorithm in Fig. 8. The unitary gate $U_n$ is the braid representation of the relevant knot, and the Jones polynomial at the fifth root of unity is approximated using the weighted trace of $U_n$, a block diagonal matrix in the computational basis. The exact structure of the relevant matrices $U_n$ (as $n$ grows) makes the problem complete for DQC1.

The algorithm requires two pseudo-pure qubits, so the target system is prepared in the state

$$\frac{1}{2^n}[2\epsilon'|00\rangle\langle 00| + (1-\epsilon')I] \otimes I^{\otimes n-1}. \tag{24}$$

---

[4] The argument regarding the number of elements to be summed is somewhat simplistic, since we only require an estimate. Nevertheless, there is good reason to expect the computation scale exponentially with n [78].

[5] Strictly speaking it should also be scalable.

As noted above, this does not change the complexity class. The matrix $U_n$ is block diagonal and the pseudo-pure qubit on the target system is rotated so that the final result is a weighed sum of the trace of the two blocks. The experiment [76] was performed on trans-chrotonic acid, a 4-qubit molecule prepared in an initial state of two pseudo-pure qubits and two maximally mixed qubits. The aim was to distinguish between six distinct Jones polynomials corresponding to braid representations encoded in the two blocks of the $8 \times 8$ unitary matrix.

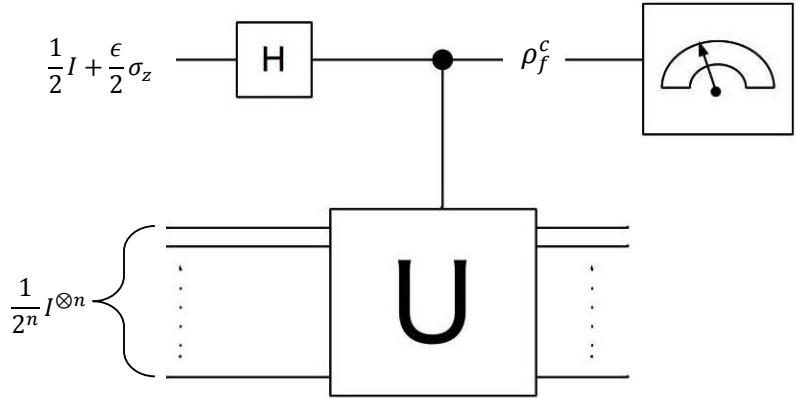

Fig. 8 The DQC1 circuit for evaluating the normalized trace of a unitary matrix with $2^n$ diagonal elements. The output state shown by Eq. (23) on the control qubit is measured to estimate the normalized trace of U. This task is believed to be computationally hard for a classical computer.

Decoherence was a major source of noise in the experiment. To compensate, the reference spectrum was measured using a similar experiment with the identity instead of the controlled unitary. The other sources of error were gate fidelities, these were maximized using GRAPE [17]. The algorithm was implemented for 18 different braids that had to be divided into six different groups corresponding to the six distinct values of the Jones polynomials. The results gave a 91% success rate at distinguishing different values.

While this experiment demonstrates good control, it is not clear how far it is possible to scale the number of qubits with similar gate fidelities. The issue of noise in DQC1 is problematic, since there is no known error correction procedure within this scheme. It is not yet known at which point this approximation algorithm fails due to errors.

The DQC1 model undermines the claim that entanglement is the essence from which quantum computer derive their power. It is known that for any bipartite cut the DQC1 algorithm can generate only a small amount of entanglement [78]. Perhaps more importantly, in the trace estimation algorithm the control is never entangled with the target. Since entanglement seems to play a major role in pure state QIP, it was suggested [74,78] that a more general form of quantum correlation called quantum discord [80,81] plays a similar role to in DQC1. Studies on the relation between quantum discord and quantum computing were extended to other measures with some success [80]. One issue with these quantities is that they are usually hard to calculate. However, a DQC1 algorithm can be used for calculating one quantum correlation measure called the geometric discord [82]. An experiment for measuring the geometric discord in DQC1 was also implemented in NMR using similar methods to the experiment for estimating the Jones polynomial [82].

## 2.4 Foundation

One of the first suggested uses for quantum computers was to test the foundations of quantum mechanics [84]. One may say that fault-tolerant quantum computers will be the ultimate test of the theory, but for the time being simpler experiments have been carried out on small quantum processors. In NMR such tests are sometimes problematic due to the major downsides of ensemble QIP: the noise and the lack of projective measurements. Nevertheless a number of experiments related to foundations have been carried out on NMR processors. In the following we describe two experiments, a state-independent test of contextuality that avoids the need for pure states and a weak measurement protocol that overcomes the inability to perform projective measurements.

*Quantum measurement and the von Neumann scheme*

Both of the protocols below rely on a particular implementation of the von Neumann measurement scheme. For a Pauli observable $\sigma_{\hat{n}}$ and a spin ½ system S initially in the state $\alpha|\uparrow\rangle + \beta|\downarrow\rangle$ (written in the eigenbasis of $\sigma_{\hat{n}}$), a projective measurement has the following properties: (a) The outcome is $\pm 1$ with probability given by the Born rule $P(+1) = |\alpha|^2$, $P(-1) = |\beta|^2$ and (b) the state of S after the measurement is the corresponding eigenstate of $\sigma_{\hat{n}}$.

In the von Neumann scheme, the measurement result is recorded on an ancillary system called the meter, in our case a spin ½ initially prepared in the $|0\rangle$ state. The measurement is a unitary interaction between the meter and the system such that, after the interaction, the state is $\alpha|\uparrow\rangle|0\rangle + \beta|\downarrow\rangle|1\rangle$. We can then say that a meter readout of $|0\rangle$ corresponds to a +1 or ↑ result and $|1\rangle$ corresponds to a -1 or ↓ result. In NMR, the meter can interact with single quantum system and then we take the average of this meter.

*Testing contextuality*

One difference between quantum and classical systems is that measurements are inherently probabilistic and disturbing. Since the early days of quantum theory there were suggestions that the probabilistic nature of the theory is due to an incomplete description and that an underlying ontological hidden variable can be used to make the theory deterministic. Theoretical results give bounds often called no-go theorems regarding the possibility of an underlying hidden variable theory, the most famous of these is Bell's theorem [85] regarding local realism. Another theorem often attributed to Bell is the Kochen-Specker theorem regarding contextuality [86]. Simply stated, the theorem shows that a hidden variable theory cannot give probabilities to measurement results independently of the context of these measurements.

Contextuality inequalities involve degenerate observables and must therefore apply to systems with at least three Hilbert space dimensions. They usually involve specific states that violate the inequality, usually pure states. Cabello [87] came up with a simple inequality which is state independent and therefore more natural for NMR. The inequality involves measurements on two spins. There are nine observables (the entries in Table 1) and six correlation measurements (the rows and columns of Table 1). Each measurement is a correlation measurement of the product of three commuting (or co-measureable) observables $O_1, O_2, O_3$ (*e.g.* $R_1 = IZ \cdot ZI \cdot ZZ$). Now let us assume that quantum mechanics is an incomplete theory and that there is a more complete description of nature (a hidden variable) such that we know that the outcomes of independent measurements of these observables should be $o_1, o_2, o_3$. Since the observables commute we know that the result of a measurement of the product $O_1 O_2 O_3$ must give the outcome $o_1 o_2 o_3$. Looking at Table 1 we can see that these products (for each row or column) are proportional to the identity, so regardless of the state they should give +1 for all row measurement and the two column measurements $c_1, c_2$ and -1 for the final column $c_3$. Adding these up we get $\beta = r_1 + r_2 + r_3 + c_1 + c_2 - c_3 = 6$. However if we try to give values to each measurement (*i.e.* each observable in the table) we cannot possibly reach the value of 6. The upper bound is 4. So the inequality reads $\beta = r_1 + r_2 + r_3 + c_1 + c_2 - c_3 \leq 4$ for all non-contextual hidden variable theories.

|       | $c_1$ | $c_2$ | $c_3$ |    |
|-------|-------|-------|-------|----|
| $r_1$ | IZ    | ZI    | ZZ    | +1 |
| $r_2$ | XI    | IX    | XX    | +1 |
| $r_3$ | XZ    | ZX    | YY    | +1 |
|       | +1    | +1    | -1    |    |

Table 1 The six measurements and nine observables for the test of contextually. For any quantum states the measurements yield the same, deterministic result. However if one tries to assign a deterministic value to each of the 9 observables the table cannot be completed. A simple analysis shows that if one assigns values to each observable the measurements obey an inequality $\beta = r_1+r_2+r_3+c_1+c_2-c_3 \leq 4$, however for any quantum state we get $\beta=6$.

An experimental test of these inequalities requires 6 correlation measurements (the rows and columns of Table 1). In NMR these correspond to different experimental setups. Each measurement is a set of three unitary interactions with a single meter spin. Each single interaction corresponds to the von Neumann scheme presented above, however since the spin ½ meter is modular i.e. two $\pi$ rotations are equivalent to no rotation, the meter only records the correlations. A $+1$ result on the meter corresponds to an even number of $+1$ results for the three obervables and a $-1$ result corresponds to an odd number. The readout is an ensemble average. Since the experiment can be done with any initial system state, including the maximally mixed state, the model requires only that the meter spin is pseudo-pure. It is therefore within the class DQC1 (see Section 3.3).

The experiment [88] was performed using a macroscopic single crystal of malonic acid with ~3% of the molecules triply abled with $^{13}$C. The gates were generated using GRAPE with an average fidelity of 99.8% and time of 1.5ms. Following the six experiments, a value of $\beta = 5.2 \pm 0.1$ was obtained, giving a violation of more than 25% with respect to the maximal classical value. Deviations from the predicted value, $\beta = 6$ can be explained by taking decoherence into account.

*Weak measurements*

Weak measurements are performed by taking the standard von Neumann measurement and making the interaction very weak. The result of such a measurement is a *shift* of the meter's wave function by a value proportional to the expectation value of the system $S$. An interesting phenomenon occurs if after the weak measurement, $S$ is measured in a different basis. In this case the shift is proportional to a weak value $\langle\phi|\sigma_{\hat{n}}|\psi\rangle/\langle\phi|\psi\rangle|$, where $|\psi\rangle$ is the preparation, $|\phi\rangle$ is the post-selection (*i.e.* the state corresponding to the result of the final measurement) and $\sigma_{\hat{n}}$ is the weakly measured observable. For non-trivial post selection the weak value can be complex and arbitrarily large.

The challenge in NMR is post-selection. Since there are no projective measurements it is impossible to post-select those molecules that gave the desired result for the final measurement. However by using the von Neumann procedure for post-selecting, and furthermore getting rid of the systems that fail post-selection by adding noise, it is possible to perform the full weak measurement [89].

Large and complex weak values are a signature of non-trivial post-selection. In an experiment [89] the final signal is diminished by $|\langle\phi|\psi\rangle|$ due to the noise added in the post-selection step. This makes very large weak values hard to observe. However, complex weak values are not difficult to observe as long as $|\langle\phi|\psi\rangle|$ is not too small. In addition decoherence causes the calculated weak values to be slightly lower than the ideal result, this is more apparent as $|\langle\phi|\psi\rangle|$ grows. Overall in the experiment weak values of 2.3 (compared with a maximal eigenvalue of 1) as well as complex weak values of magnitude 1 were observed with good precision. Attempts to observe larger weak values did not reach the theoretical predictions due to decoherence and relatively strong coupling that had to be used to counter the small signal.

## 2.5 Quantum simulation

Simulating a generic quantum system is believed to be a hard problem for a classical computer. Due to the exponential size of the Hilbert space as a function of subsystems a simulation may require an exponential number of parameters that need to be stored and updated. In 1982 Feynman suggested that a controllable quantum system can be used to simulate other quantum systems [90]. Feynman's idea was one of the motivating forces for quantum computing. However in many cases quantum simulators are easier to construct than universal quantum computers. In the past decades, progress has been made in making the first steps in using one quantum system to simulate another one [91]. In the following we outline a few quantum simulation experiments done in NMR.

*Digital quantum simulation of the statistical mechanics of a frustrated magnet* [92]

The straightforward approach to simulate the evolution of a quantum system is by directly implementing a similar Hamiltonian on a different system. This method often called *analog* [93] requires the design of a very specific physical system that can simulate a very specific set of Hamiltonians. In *digital* quantum simulations, the initial state is represented by qubits and the time evolution is approximated by applying a sequence of short-time unitary gates [94]. The digital simulators are more general and usually require better control, often to the level of universal quantum computers.

To study the ground state of an Ising Hamiltonian, it is possible to use an adiabatic (analogue) method, however this method requires that the energy gap between the ground state and first excited state is large enough to avoid excitations. In general it is very hard to determine energy gap efficiently [95]. In the digital simulation [88] the groud state is prepared using a quantum circuit which is composed efficently from the Hamiltonian.

An Ising Hamiltonian for 3 spin ½ particles is given by [96]
$$H = J(Z_1 Z_2 + Z_2 Z_3 + Z_1 Z_3) + h(Z_1 + Z_2 + Z_3), \qquad (25)$$
where $Z_i$ is the Pauli-Z matrix for the spin $i$. For $J > 0$, the coupling is anti-ferromagnetic, and the spins tends to align in opposite direction to minimize energy. On the other hand, $h$, has the effect of aligning spins in same direction. These two opposite forces result in a frustrated ground states and a very rich phase diagram.

At $h = 0$, there are six possible configuration of ground state possible i.e. the ground state is six-fold degenerate, so the entropy of the system can be non-zero even at $T \to 0$. Now if by changing $h$ a little bit in either positive or negative direction, three of the six configurations are preferred. By increasing $h$ further such that it becomes the dominant factor in the Hamiltonian, all the spins will align in the same direction and the ground state will be non-degenerate at $h = -2J$ and $2J$, see Fig. 9. The purpose of the simulation was to study the magnetization, $Z_1 + Z_2 + Z_3$ and entropy equals to $\text{Tr}(\rho \log \rho)$, where $\rho$ is the density matrix of Ising spin chain of the three spin system as a function of $h$.

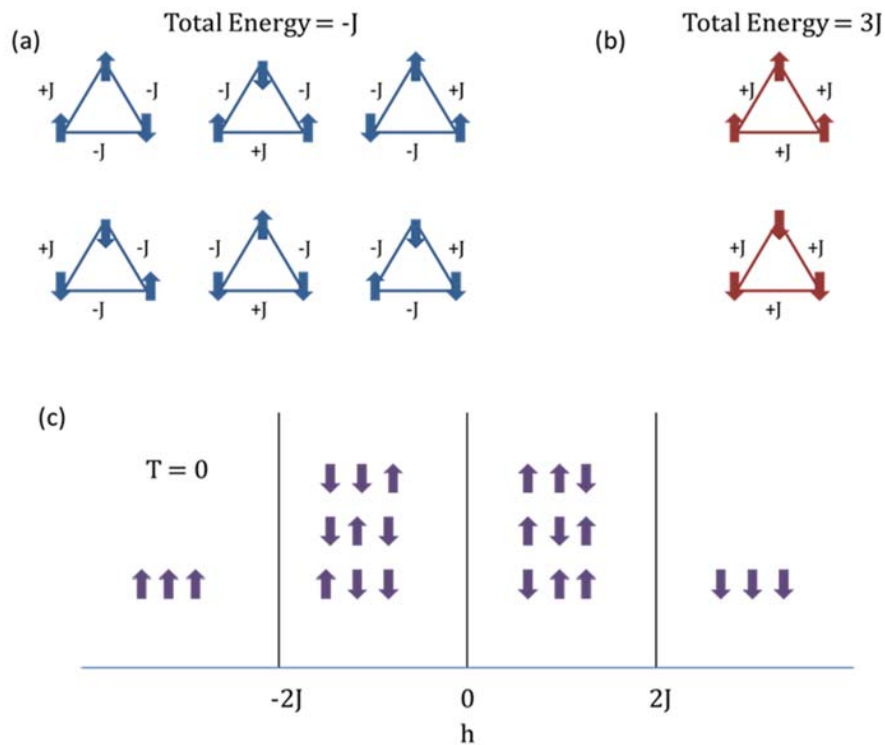

Fig. 9 (a) The six possible configuration for the ground state of Ising spin chain with a magnetic field h=0, (b) the configuration at $h=\pm 2J$, and (c) The configurations preferred as a function of $h$.

The total magnetization as a function of $h$ is a step function. For $h < -2J$ the spins all the spins will point upwards, giving maximum magnetization in upward direction. For $-2J < h < 0$, the tendency to align anti-parallel due to the first part of the Hamiltonian makes one out of three spins to point downwards, similarly, for $0 < h < 2J$, two of three spins face downwards and for $h > 2J$, all of them point downwards. For $h = 0$, all 6 combinations are equally likely, the system has zero net magnetization.

As for entropy: In the region $h < -2J$ and $h > 2J$ all the spins are aligned in the same direction the entropy is minimal. , similarly in $-2J < h < 2J$ (excluding the point $h = 0$) the ground state is threefold degenerate and the entropy is higher. The maximum entropy being at $h = 0$. Two more points with higher entropy are $h = -2J$ and $h = 2J$.

In the experiment $^{13}C$- labeled trans-crotonic acid was used as the four qubit processor. Three carbons were used to simulate the three spin chain while the fourth one was used to measure the expectation value of various Pauli operators at the end of the protocol and reconstruct the density matrix. This way of reading off elements from the fourth carbon's spectra was useful due to its well resolved spectra compared to the other three. The starting state was a coherent encoding of the thermal states [97,98]. Different unitary evolutions corresponded to variation in the Hamiltonian parameters. These were decomposed into single-qubit gates and 2-qubit evolutions so that the variation in the Hamiltonian parameters required only phase shift in the applied RF pulses. This method exploits the ability to manipulate phases in the RF pulses to high precision (resolution of $10^4$ per rad). Since entropy is a non-linear function of the state, small errors in the density matrix result in higher errors in the calculated entropy.

*Quantum simulation of entanglement in many body systems* [80]

At zero temperature all the thermal fluctuations resulting in phase transition cease, but as seen in the previous experiment, quantum fluctuation can still occur. These quantum fluctuations give rise to quantum phase transitions (QPTs), under specific conditions. These conditions can be tuned by controlling a parameter in the Hamiltonian of the system, for example, the magnetic field. The result of these quantum fluctuations is an abrupt change in the ground state wave function [100]. It is not always possible to keep track of the ground state wave function to see when the QPT has occurred, but one can use different properties which are easier to track. One quantity that plays a role is entanglement, however, measuring entanglement in a many-body system is a challenging task [101,102]. In Ref. [80] the ground state of an *XXZ* spin chain [103] was simulated and the geometric entanglement (GE) was measured to study QPTs.

The GE of a pure state $|\Phi\rangle$ is given by the expression [104,105]:

$$E|\Phi\rangle = -\log_2 \Lambda_{max}^2 \quad , \text{with} \quad \Lambda_{max} = \max_{\Psi} |\langle\Psi|\Phi\rangle| \qquad (26)$$

Where maximization is over all pure product states. $\Lambda_{max}$ can be interpreted as the distance from the closest product state $|\Psi\rangle$ which is equivalent to the probability to end up in this state after performing the optimal local projective measurement on every spin. Experimentally GE can be obtained by an iterative method, which converges very quickly to its maximal value as a function of number of iterations.

The iterative method is: choose a random local measurement basis and pick one direction for each spin. Vary the basis of the $i^{th}$ spin, to find the measurement that yields the largest measurement probability. Next, move on to the next, $(i+1)^{th}$ spin and perform the same procedure. Repeat the procedure by sweeping back and forth until the global maximum is reached.

The *XXZ* spin chain is described by the Hamiltonian

$$H_{XXZ} = \sum_{i=1}^{N} (X_i X_{i+1} + Y_i Y_{i+1} + \gamma Z_i Z_{i+1}), \tag{27}$$

Where $X_i$, $Y_i$, and $Z_i$ denotes the Pauli matrices acting on $i^{th}$ spin, $\gamma$ is a control parameter and the boundary condition are periodic i.e. $N + 1 \equiv 1$. The *XXZ* chain can be solved exactly [103] and its ground states have a rich phase diagram. For $\gamma < -1$, the ground state is in a ferromagnetic Ising phase. At $\gamma = -1$, first order phase transition occurs and from $-1 < \gamma \leq 1$ the ground state is in a gapless phase. Again at $\gamma = 1$ there is an $\infty$ order phase transition and for $\gamma > 1$ the ground state is in the anti-ferromagnetic phase [106].

The calculation of ground state energy shows discontinuity at $\gamma = -1$, whereas it is analytic across $\gamma = 1$ and so is any correlation function. As a consequence measures based on correlation functions are insensitive to the $\infty$ order transition. GE on contrast shows a jump at $\gamma = -1$ and a cusp (i.e. the derivative is discontinuous) at $\gamma = -1$, making it a better measure to study QPTs [107].

In principle the closest product state is not known and the iterative method described above provides an efficient method to find the maximum overlap and hence the GE. The advantage of measuring GE was that both first and $\infty$ order quantum phase transition were detected, with the transition points being when either GE or its first order derivative is discontinuous. On the other hand, all the traditional statistical-physical methods for QPTs, such as correlation functions and low-lying excitation spectra will be continuous at the $\infty$ order phase transition, since the ground state energy is continuous.

The experiment was performed using the four carbons in crotonic acid dissolved in acetone-d6. Various ground states could be prepared by only varying single spin rotations and the quantum gates were prepared using the GRAPE algorithm [17]. For $\gamma < -1$, the experimental value of $\Lambda_i^2$ was measured to be 0.92, 0.048, and 0.019 compared to theoretical values of 1, 1/16, and 0 respectively. A polynomial fit of $\Lambda_2^2$ and $\Lambda_3^2$ shows that they intersect at $\gamma = 0.92$ compared with the theoretical value, $\gamma = 1$. To account for this discrepancy an inverse decay parameter was added to experimental data. The new experimental value decay into account gives an intersection of $\Lambda_2^2$ and $\Lambda_3^2$ at $\gamma = 1.02$.

*Quantum data bus in dipolar coupled nuclear spin qubits* [108]

In both (quantum and classical) forms of computation and communication one important task is to transfer an arbitrary state from one (qu)bit to another. Many quantum state transfer (QST) protocols exists [109–112] such as applying a SWAP gate between the qubits. The SWAP gate can be applied by evolving the qubits of interest under the dipolar coupling, but experimentally it is difficult when the spins cannot be addressed individually, for example, in large-size solid state systems. In Ref. [108] a QST protocol was implemented by exploiting a scheme of applying gates iteratively to only two qubits at one end of the qubit chain, irrespective of the size of the chain. Each iteration transfers part of the information from the point of origin to the desired location in the chain, and with fidelity of transfer reaching unity with more number of iteration performed. Dipolar couplings of a liquid crystal sample were used. These are much stronger than the scalar couplings, making gate time significantly shorter [113]. The benefit of this scheme is that one does not need a global control of the spin, or individual control of all the spins in the chain. Two individually addressable qubits are sufficient to perform this protocol.

The aim is to transfer an arbitrary state $\alpha|0\rangle + \beta|1\rangle$ from $j$ to $N$ in a $N$-spin system. The Hamiltonian for the spins up to $N-1$ is given by

$$H = \frac{1}{2}\pi \sum_{(j,k=1;k>j)}^{(N-1)} D_{jk}(2\sigma_z^j\sigma_z^k - \sigma_x^j\sigma_x^k - \sigma_y^j\sigma_y^k) \qquad (28)$$

where $\sigma_{x,y,z}^j$ represents the Pauli matrices with $j$ representing the spin on which it acts. Evolving the N-1 spins under this Hamiltonian for time $\tau$ gives $U_\tau = e^{-i\tau H}$ with

$$U_\tau|\mathbf{0}\rangle = e^{i\theta}|\mathbf{0}\rangle \ ; \qquad U_\tau|\mathbf{j}\rangle = \sum_{k=1}^{N-1} a_k\ |\mathbf{k}\rangle \qquad (29)$$

where $|\mathbf{0}\rangle\ and\ |\mathbf{j}\rangle$ represent, all spin pointing up and all spins up except the spin $j$

pointing down respectively. The bold characters indicate multiple spin state. Now the main iterative gate is carefully chosen such that, after $n$ iterations the following transformations occur $(\alpha|0\rangle + \beta|j\rangle) \rightarrow \alpha e^{in\theta}|0\rangle + \beta|N\rangle$, where $e^{in\theta}$ is a known phase induced by the gate $U_\tau$. Since the process is unitary we can invert it to transfers arbitrary state from position N to j. Hence one can prepare an arbitrary state at any location of the spin chain by having control on only two spins of the chain. Another useful extension of this method is to be able to prepare an entangled state between any two locations of the spin chain. This can be easily achieved by slightly changing the way in which the iterative gate is calculated.

The experiments were performed on a 4 spin chain provided by the four protons of the orthochlorobromobenzene (C6H4ClBr) dissolved in a liquid crystal solvent ZLI-1132. We transferred $\sigma_x, \sigma_y,$ and $\sigma_z$ from spin 1 to spin 4 on the four spin system we have, the experimental fidelities were $0.654 \pm 0.046, 0.660 \pm 0.052,$ and $0.693 \pm 0.037$ respectively after 100 iterations. The next part of experiment involved entangled state where spin 1 and 4 are entangled. The experimental fidelity was calculated as 0.77. The major sources of imperfections are attributed to the inhomogeneities of the magnetic field, imperfect implementation of the GRAPE pulses, and to the decoherence.

## 3. Conclusion and Perspective

Liquid state NMR, one of the first proposals for quantum processors [11,12], has clearly demonstrated major steps towards the realization of ideas and concepts of quantum information science in the laboratory. However, NMR is an unlikely candidate for a quantum computer due to the lack of scalability in practice. Despite impressive control, the ratio of gate time to decoherence is still too small as the sizes of systems grow to more than a dozen qubits. One way to address this limitation is to extend liquid-state NMR to solid-state NMR, where various dynamical nuclear polarization techniques can be employed and the speed of gate operations can be increased via much larger dipolar couplings. Another way to achieve scalability is to involve electrons as actuators in electron spin resonance (ESR) systems [114], to achieve indirect control of nuclear spins in a much faster approach [115].

Even if other platforms will be used to implement an eventual quantum computer, NMR still plays a leading role in progressing towards this goal. The experiments and techniques reviewed in this chapter should convince the reader that most quantum computing schemes within seven qubits or less are reasonably straightforward to implement in NMR. The control demonstrated in NMR exceeds the capabilities of any other system used today. The advanced techniques developed in NMR quantum computation, such as GRAPE pulses and pulse fixing, have been extended to many other systems successfully to realize high-fidelity control. The lessons learned in the history of NMR quantum computation have and continue to be indispensable in the development of experimental quantum computation.

**Acknowledgement**

We thank Rolf Horn for helpful comments and discussions. This work is supported by Industry Canada, NSERC and CIFAR.